\documentclass[twocolumn,twoside,slac_two]{revtex4}
\usepackage{graphicx}
\usepackage{fancyhdr}
\begin{document}

\title{ATLAS Muon Detector Commissioning}

%

\author{E. Diehl, for the ATLAS muon collaboration}
\affiliation{Department of Physics, University of Michigan, Ann Arbor,
  MI 48103, USA}

\begin{abstract}
The ATLAS muon spectrometer consists of several major components:
Monitored Drift Tubes (MDTs) for precision measurements in the bending
plane of the muons, supplemented by Cathode Strip Chambers (CSC) in
the high $\eta$ region; Resistive Plate Chambers (RPCs) and Thin Gap
Chambers (TGCs) for trigger and second coordinate measurement in the
barrel and endcap regions, respectively; an optical alignment system
to track the relative positions of all chambers; and, finally, the
world's largest air-core magnetic toroid system. We will describe the
status and commissioning of the muon system with cosmic rays and plans
for commissioning with early beams.
\end{abstract}

\maketitle

\thispagestyle{fancy}


\section{Introduction}

The ATLAS experiment is one of two general-purpose collider detectors
for the Large Hadron Collider (LHC) at CERN.  The ATLAS detector
consists of an inner detector employing silicon pixel, strip, and
transition radiation tracking detectors, all in a solenoidal magnetic
field of 2 Tesla; electromagnetic and hadronic calorimeters using
liquid argon and scintillator tile detectors; and a muon spectrometer.
The muon spectrometer consists of a large air-core barrel and endcap
toroid magnets with a typical field of 1 Tesla, and four types of
trigger and precision tracking detectors, described below.  The
spectrometer is designed to measure the transverse momentum ($p_T$) of
muons with $p_T > 3$ GeV with a resolution of 3\% for $p_T < 250$ GeV
and increasing to 10\% @ 1 TeV.  This paper describes the
commissioning of the ATLAS muon detector for the first LHC collisions
which are expected in Fall, 2009.
  
\section{Muon Spectrometer Overview}

The ATLAS muon spectrometer consists of monitored drift tubes (MDTs)
for precision tracking in the spectrometer bending plane, Resistive
Plate Chambers (RPCs) and Thin Gap Chambers (TGCs) for triggering in
barrel and endcap, respectively, and Cathode Strip Chambers (CSCs) for
precision measurements in the high-rate endcap inner layer where MDTs
would have occupancy problems.

\begin{figure}[ht]
\centering
\includegraphics[width=80mm]{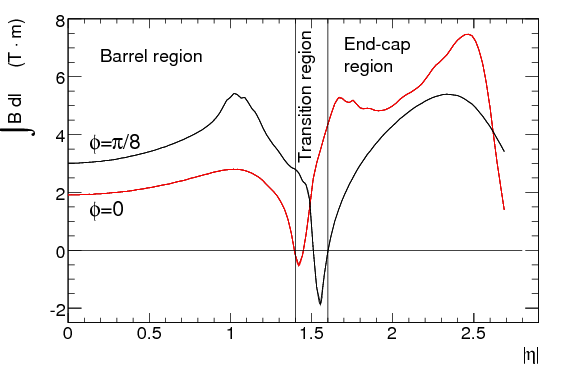}
\caption{$B \cdot dl$ of the ATLAS muon spectrometer.} 
\label{fig:magnetic_field}
\end{figure}

The magnet system consists of 3 sets of air-core toroids, each with 8
coils, 1 for the barrel, and 1 for each endcap.  The barrel toroids
coils are each 25m $\times$ 7m and the endcap coils are 9m $\times$
4m.  The magnetic field provides an approximately 1T field at the
center of each coils, but is rather non-uniform, especially in the
barrel-endcap transition region.  For track reconstruction, the field
is mapped using computer models of the field which are normalized to
measurements from 1850 Hall sensors mounted on spectrometer chambers.
Figure \ref{fig:magnetic_field} shows the $B \cdot dl$ of the
spectrometer magnetic field.

Alignment measurements of the spectrometer are also critical for
momentum determination and are accomplished with an optical alignment
system of 12k sensors.  Measurements from these sensors allow a 3
dimensional reconstruction of chamber positions accurate to better
than 50 $\mu$m.  In addition, the optical alignment system is
complemented by alignment done with tracks.

\begin{figure}[ht]
\centering
\includegraphics[width=80mm]{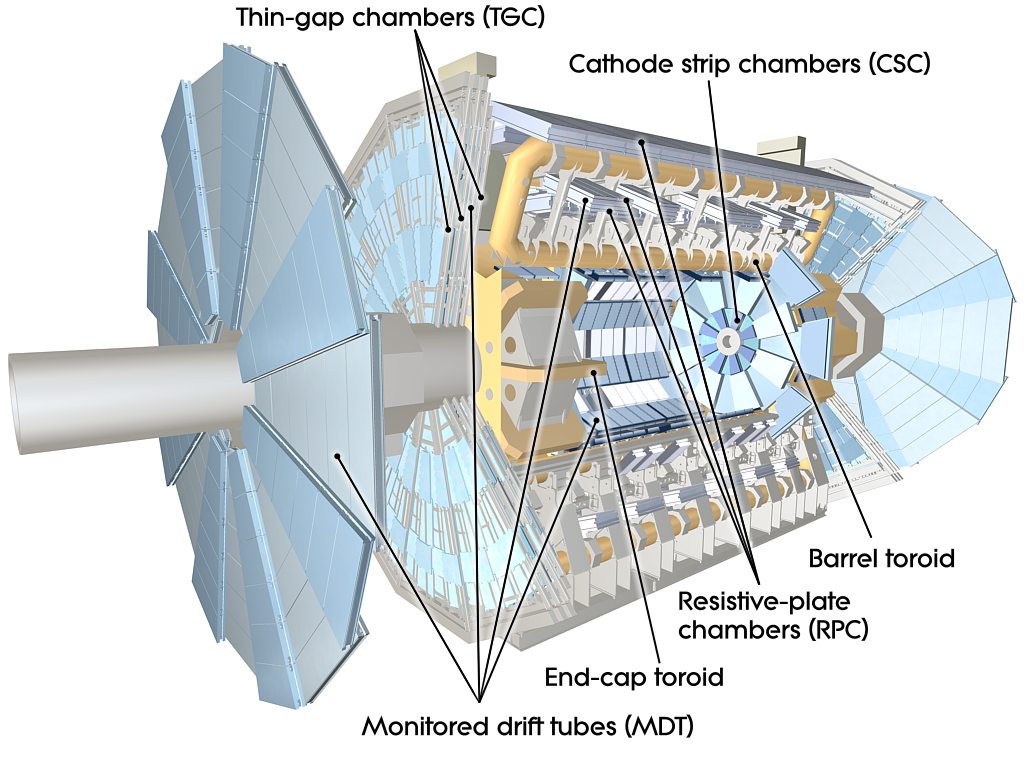}
\caption{Layout of the muon spectrometer.} \label{fig:layout}
\end{figure}

Table \ref{table:muon_spectrometer} gives a summary of the muon
spectrometer detector components and Figure \ref{fig:layout} shows
the layout.

\begin{table}[h]
\begin{center}
\caption{ATLAS Muon Spectrometer.}
\begin{tabular}{|l|c|c|c|c|}
\hline \textbf{Type} & \textbf{Purpose} & \textbf{location} &
\textbf{$\eta$ coverage} & \textbf{Channels}  \\
\hline 
MDT & Tracking & barrel+endcap      & $0.0 < \eta < 2.7$ & 354k \\
CSC & Tracking & endcap layer 1     & $2.0 < \eta < 2.7$ & 30.7k \\
RPC & Trigger  & barrel             & $0.0 < \eta < 1.0$ & 373k \\
TGC & Trigger  & endcap             & $1.0 < \eta < 2.4$ & 318k \\
\hline
\end{tabular}
\label{table:muon_spectrometer}
\end{center}
\end{table}

The spectrometer is designed so that muons cross three layers of MDT
chambers for the sagitta measurement.  The track coordinate in the
bending plane of the spectrometer is measured by the precision
chambers with a resolution of 60-70 $\mu$m.  In comparison, the
sagitta of a 1 TeV muon will be about 500 $\mu$m.  The trigger
chambers are placed on opposite sides of the middle MDT layer.  The
trigger chambers provide a trigger based on muon momentum in addition
to identifying the bunch crossing time of the muon.  The also provide
the second coordinate measurement (non-bending plane) accurate to 5-10
cm.
 
The following sections describe the commissioning of the various
components of the muon spectrometer.  The data shown are all from
studies with cosmics rays made in 2008 and 2009.  A longer description
of the ATLAS detector has been published in \cite{atlas_paper}.


\section{RPC commissioning}

\begin{figure}[ht]
\centering
\includegraphics[width=60mm]{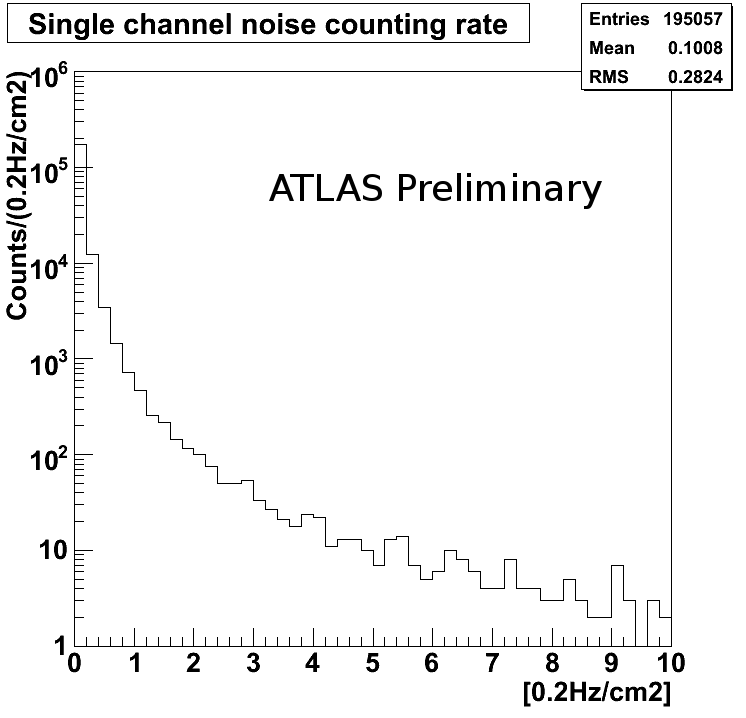}
\caption{RPC noise.  The mean noise is $\rm 0.1 Hz/cm^2$.} 
\label{fig:rpc_noise}
\end{figure}

\begin{figure}[ht]
\centering
\includegraphics[width=80mm]{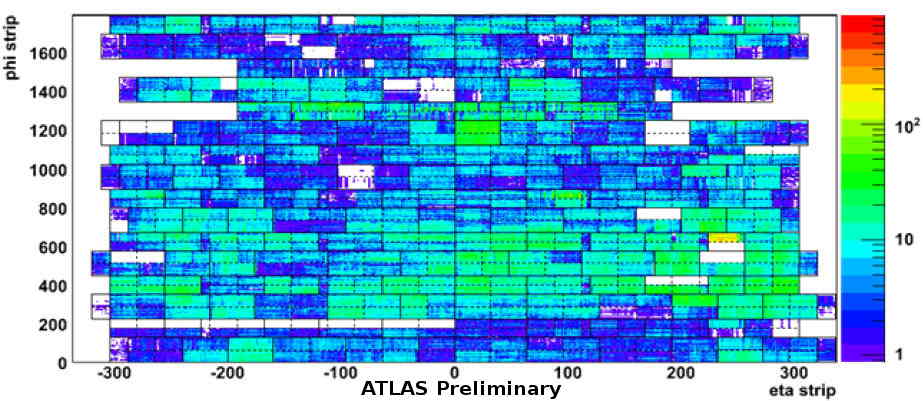}
\caption{RPC Occupancy - number of events per $\eta$ and $\phi$
  strip from a cosmic-ray commissioning run. } 
\label{fig:rpc_occupancy}
\end{figure}

\begin{figure}[ht]
\centering
\includegraphics[width=80mm]{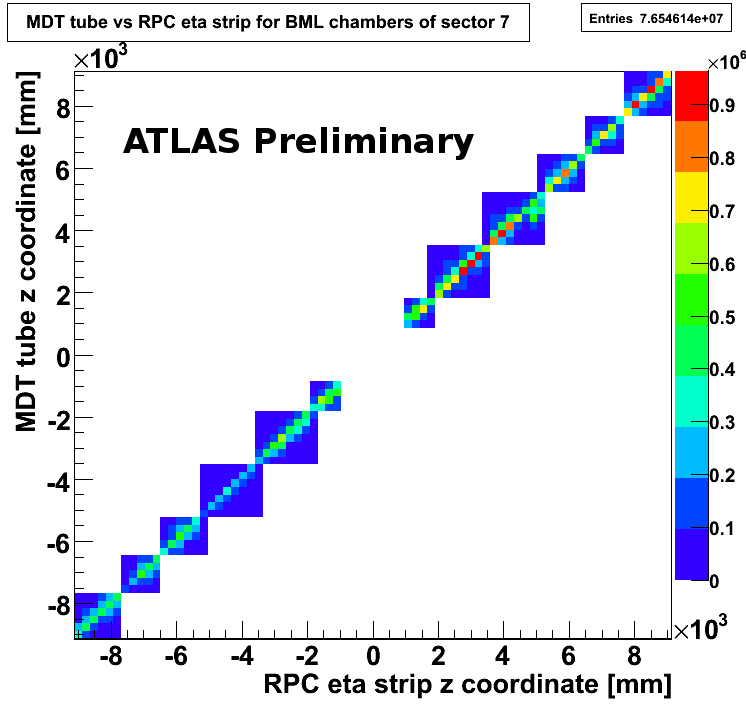}
\caption{Correlation RPC-MDT hits vs $\eta$.  The center gap is due to
  the service opening at the center of ATLAS which is not
  instrumented.  The blue background in each box is due to
  uncorrelated hits.
} \label{fig:rpc_mdt}
\end{figure}

All 606 RPC have been installed and 95.5\% are functioning as of July
2009.  Further commissioning efforts will push the percentage higher
before the first collisions.  Figure \ref{fig:rpc_noise} shows the RPC
noise measured in the ATLAS cavern.  Noise rates are typically quite
low, around $\rm 0.1 Hz/cm^2$.  Figure \ref{fig:rpc_occupancy} shows
the RPC occupancy measured in January 2009.  The occupancy is fairly
uniform, though a few chambers are missing.  Many of the missing
chambers have since been brought online.  Figure \ref{fig:rpc_mdt}
shows the correlation between RPC and MDT hits for one barrel sector.
The plot shows the expected correlation.  The blue background is due
to uncorrelated noise hits.

\section{TGC  commissioning}

\begin{figure}[ht]
\centering
\includegraphics[width=80mm]{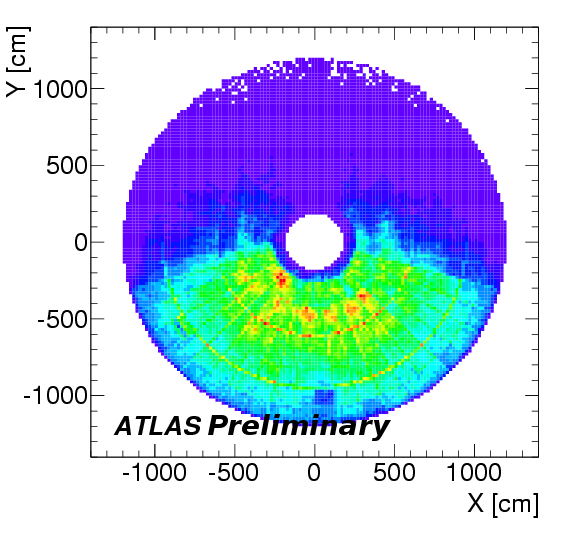}
\caption{TGC occupancy. Only the bottom of the detector was configured
for triggering in this run.} 
\label{fig:tgc_occupany}
\end{figure}

\begin{figure}[ht]
\centering
\includegraphics[width=80mm]{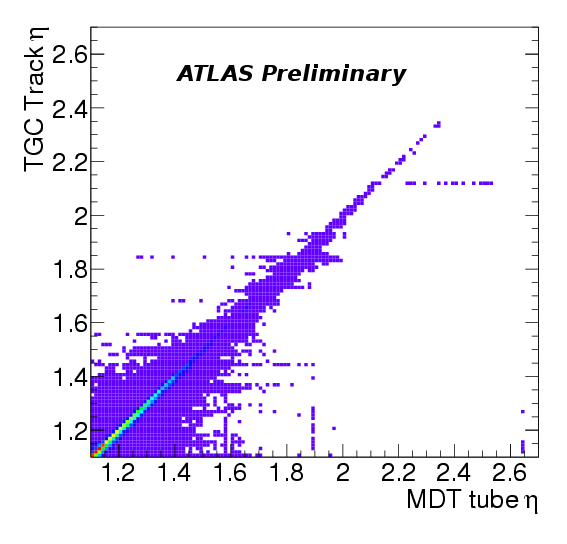}
\caption{The correlation of TGC-MDT hits vs $\eta$.  Horizontal and
  vertical lines are due to noisy chambers.}
\label{fig:tgc_mdt}
\end{figure}

\begin{figure}[ht]
\centering
\includegraphics[width=80mm]{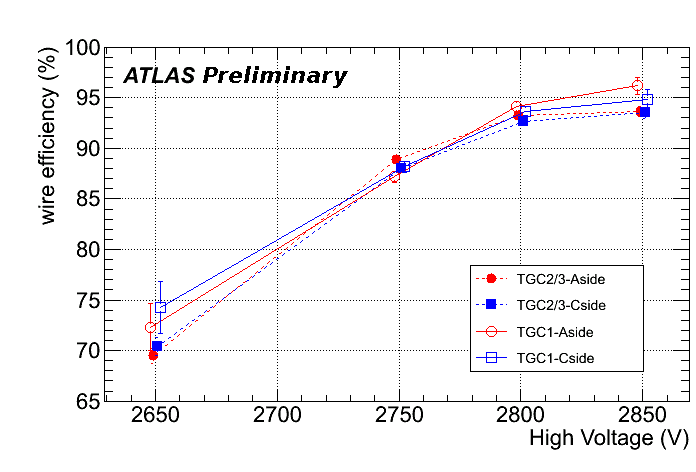}
\caption{TGC efficiency versus operating voltage.} 
\label{fig:tgc_eff_vs_voltage}
\end{figure}

All 3588 TGC chambers have been installed in the endcaps with 99.9\%
functioning.  Figure \ref{fig:tgc_occupany} shows an occupancy plot
for one of the TGC endcap wheels.  In this data only the bottom half
of the detector was set to trigger to select (cosmic-ray) tracks which
point to the interaction point, which accounts for the non-uniform
occupancy.  Figure \ref{fig:tgc_mdt} shows the correlation between TGC
and MDT in $\eta$.  The expected correlation is seen, along with a few
vertical and horizontal lines due to noisy chambers.  Figure
\ref{fig:tgc_eff_vs_voltage} shows the TGC wire efficiency versus
voltage which is around 95\% at the operating voltage of 2850V.
Trigger timing has been adjusted so that 99.14\% of events are read
out in the current (i.e. correct) bunch crossing.

\section{CSC commissioning}

\begin{figure}[ht]
\centering
\includegraphics[width=80mm]{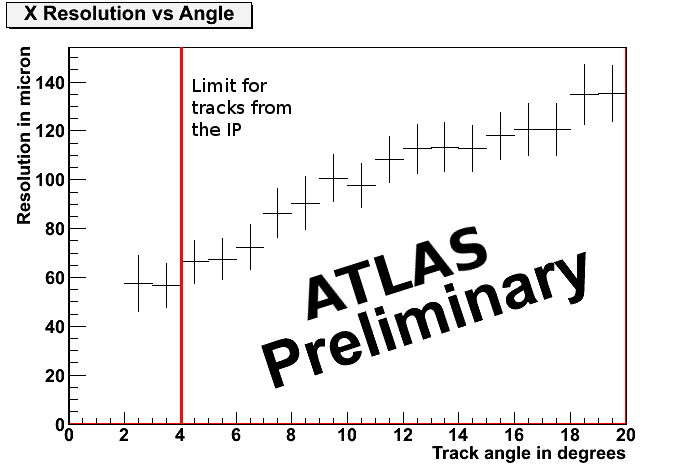}
\caption{CSC Resolution as function of relative track angle as
  measured with cosmic rays.  Tracks from interaction point will all
  be with $4^\circ$. } \label{fig:csc_res}
\end{figure}

The CSCs are installed in the inner high-$\eta$ region of the first
endcap wheel where the background will be highest.  All 32 chambers
have been installed with 98.5\% of the layers working.  There had been
a problem with poor performance of the readout firmware, but a recent
firmware re-write shows greatly improved results and we anticipate a
smoothly running system in time for the beam.  Figure
\ref{fig:csc_res} shows the CSC tracking resolution as a function of
track incidence angle using cosmic-ray data.  The resolution is around
60 $\mu$m for tracks within $4^\circ$, the angular limit for tracks
from collisions.

\section{MDT commissioning}

\begin{figure}[ht]
\centering
\includegraphics[width=80mm]{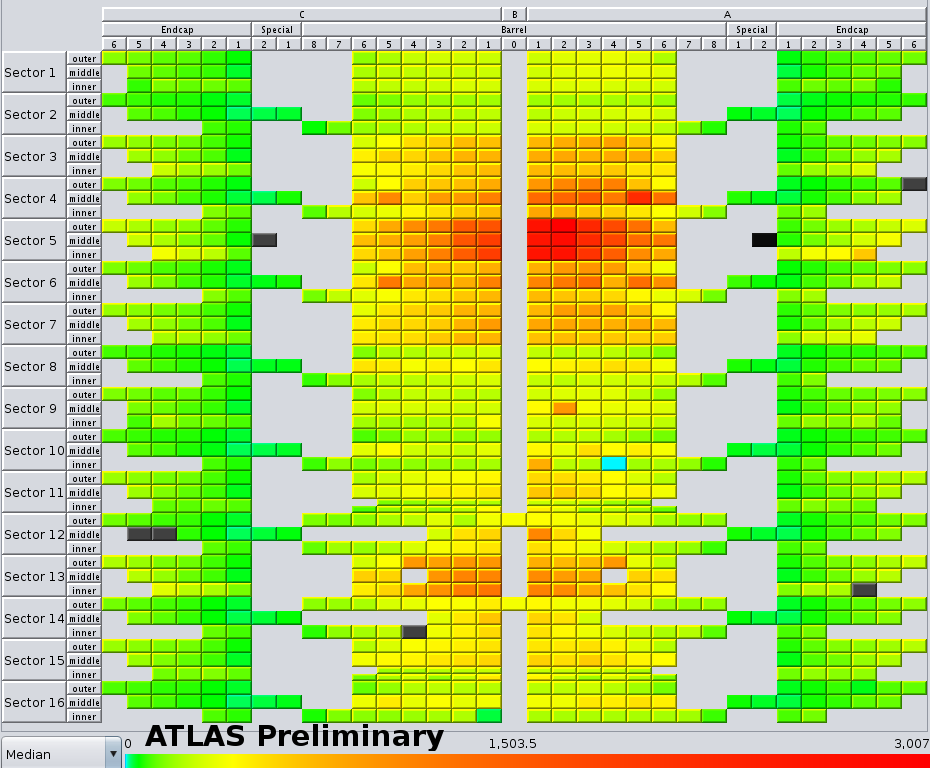}
\caption{MDT Occupancy as a function of $\eta$ and $\phi$ (sector).
  The reddish higher occupancy regions are due to the access shafts
  which allow in more cosmic rays.} \label{fig:mdt_occ}
\end{figure}

In the MDTs 1090 of 1150 chambers are installed and 99.6\% are
working.  The remaining few chambers will be installed over the coming
year.  Figure \ref{fig:mdt_occ} shows the MDT occupancy for cosmic-ray
data.

\begin{figure}[ht]
\centering
\includegraphics[width=80mm]{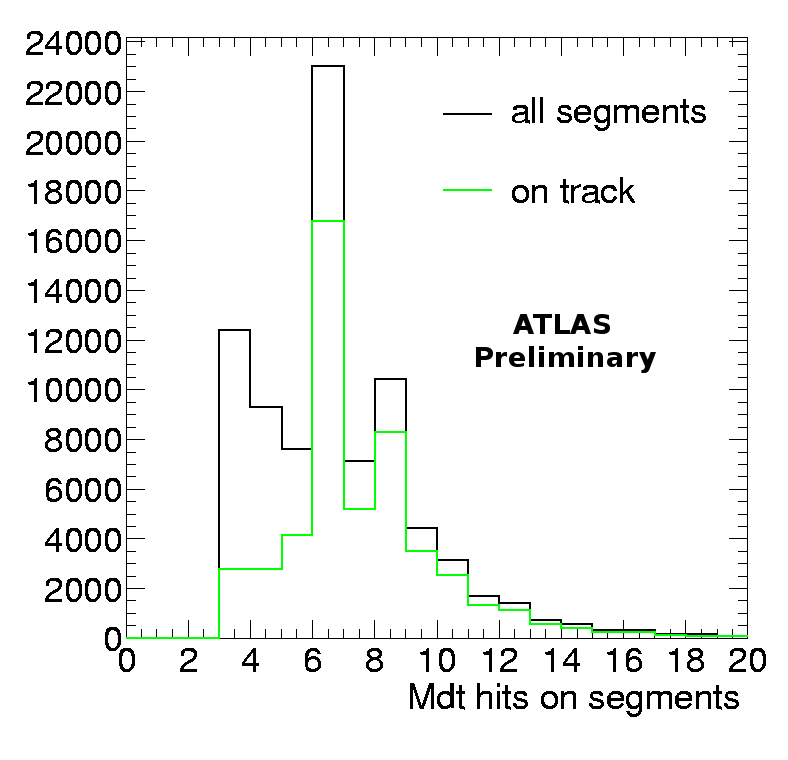}
\caption{Number of hits in MDT segments (i.e. track segment
  reconstructed within a single MDT chamber).  Complete tracks combine
  segments from 3 MDT chambers.}
\label{fig:mdt_segment_hits}
\end{figure}

\begin{figure}[ht]
\centering
\includegraphics[width=80mm]{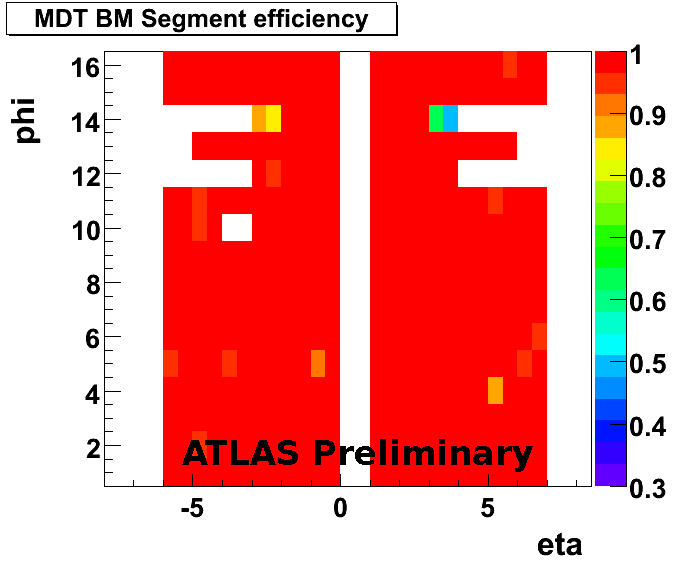}
\caption{Segment efficiency in the MDT middle barrel layer.  Cutouts
  at $\phi$=12,14 for ATLAS support structures are clearly visible.}
\label{fig:bm_segment_eff}
\end{figure}

Track reconstruction has been carefully studied in the spectrometer.
Figure \ref{fig:mdt_segment_hits} shows the distribution of the number
of hits in track segments reconstructed in MDT chambers.  The
distribution shows peaks at 6 and 8 hits which correspond to the
number of tube layers in MDT chambers.  Studies of track
reconstruction in MDTs show that the efficiency of reconstruction of
track segments within chambers is $> 98\%$.  This efficiency was
determined extrapolating tracks reconstructed from 2 chambers to a
$\rm 3^{rd}$ chamber traversed by the track and then checking for a
segment matching the incident track.  Figure \ref{fig:bm_segment_eff}
shows the segment efficiency in the MDT middle barrel layer.

MDTs require calibrations to determine a timing offset ($\rm T_0$),
and a time-to-space function (RT function).  These calibrations will
be performed with a special data stream from the level-2 trigger which
will deliver up to $10\times$ the regular rate of single muons to
provide high statistics for calibrations.  This data stream will be
processed at 3 off-site calibration centers in Rome, Munich, and
Michigan.

\begin{figure}[ht]
\centering
\includegraphics[width=80mm]{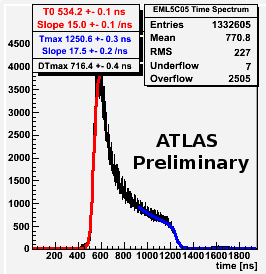}
\caption{MDT drift time spectrum with $\rm T_0$ fit shown in red.} 
\label{fig:tzero_fit}
\end{figure}

\begin{figure}[ht]
\centering
\includegraphics[width=80mm]{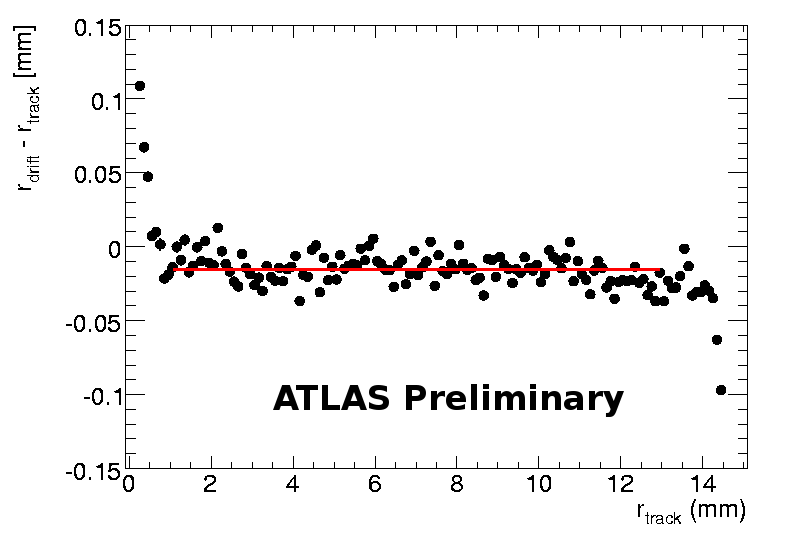}
\caption{MDT tracking residuals vs tube radius after autocalibration.
  Distribution is flat and narrow.}
\label{fig:residuals_radius}
\end{figure}

The $\rm T_0$'s are determined by fitting the rising edge of the drift
time spectrum with a Fermi-Dirac function as shown in Figure
\ref{fig:tzero_fit}.  The RT function is determined by an
autocalibration technique minimizing track residuals using either
chamber data, or data from an MDT monitoring chamber on the surface
which samples the MDT gas.  Figure \ref{fig:residuals_radius} shows an
example of tracking residuals versus tube radius after RT
auto-calibration.  Additional details about the ATLAS MDT calibration
process are discussed in more detail in the DPF talk of S. McKee
(``ATLAS Great Lakes Tier-2 Computing and Muon Calibration Center
Commissioning'') and in \cite{calib_paper}.

\begin{figure}[ht]
\centering
\includegraphics[width=80mm]{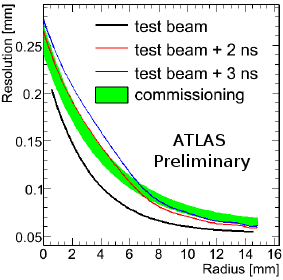}
\caption{MDT resolution for cosmic-ray and testbeam data.  See text
  for explanation.} 
\label{fig:mdt_res}
\end{figure}

Figure \ref{fig:mdt_res} shows the MDT resolution versus tube radius
as measured with both cosmic rays and testbeam
data\cite{testbeam_paper}.  The resolution obtained with cosmic-ray
data is poorer than that measured in testbeam data.  The cosmic-ray
resolution could be matched roughly by adding an additional jitter of
about 2 ns to testbeam plot.  The decrease in resolution for cosmic
rays is due to the non-synchronous arrival of cosmic rays compared to
the 25 ns beam clock which is used for clocking the drift time
measurements.  This cosmic-ray jitter is mostly removed with an
event-by-event $T_0$ fitting procedure, but a residual jitter remains.
We expect that with collision data we will be able to match the
testbeam resolution.

\section{Alignment system commissioning}

\begin{figure}[ht]
\centering
\includegraphics[width=60mm]{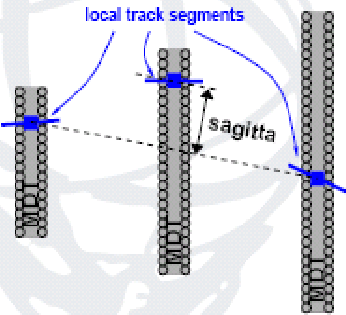}
\caption{Sagitta definition for alignment studies.} 
\label{fig:sagitta_diagram}
\end{figure}

\begin{figure}[ht]
\centering
\includegraphics[width=80mm]{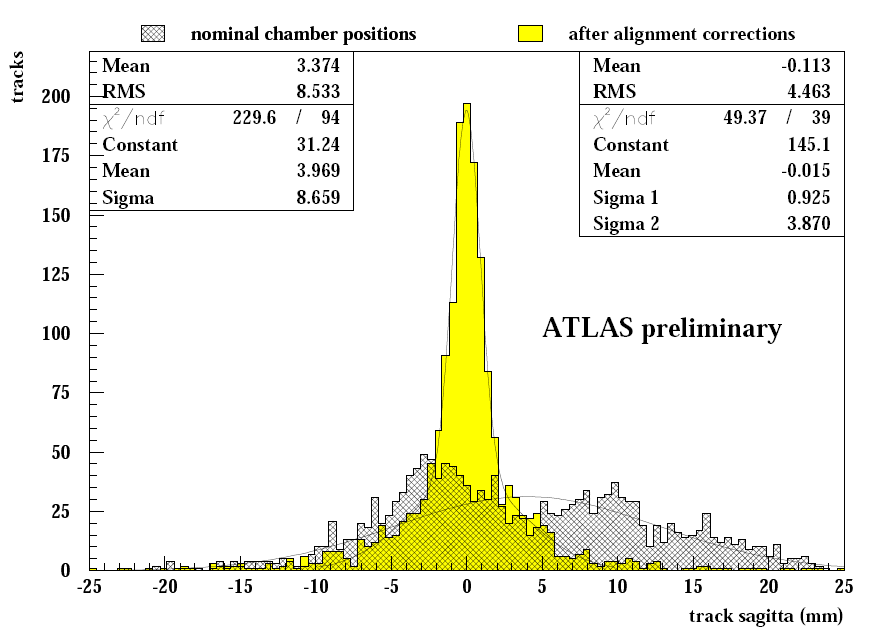}
\caption{Sagitta in endcap cosmic rays with and without optical alignment
  corrections.} \label{fig:endcap_align}
\end{figure}

The ATLAS muon spectrometer has separate optical alignment systems for
the barrel and endcap.  These employ a network of sensors to provide a
3D reconstruction of detector positions, which are essential for the
muon sagitta measurement, and therefore momentum measurement of muons.
The endcap optical system has been validated by looking at the
``false'' sagitta of straight tracks (i.e. with no magnetic field)
which pass through the 3 endcap wheels.  A track is reconstructed by
hits on the inner and outer wheels and a sagitta calculated relative
to this track using hits from the middle wheel as shown in Figure
\ref{fig:sagitta_diagram}.  Figure \ref{fig:endcap_align} shows the
sagitta distribution with and without the alignment corrections
applied.  Without corrections, using nominal geometry, the sagitta
distribution is broad and distorted.  When alignment corrections are
applied the distribution is centered on zero to within 15 $\mu$m,
(i.e. no net sagitta, as expected for straight tracks) with a much
narrower and uniform shape.  The width of the distribution is
primarily due to multiple scattering which is relatively high in
cosmic rays.

\begin{figure}[ht]
\centering
\includegraphics[width=70mm]{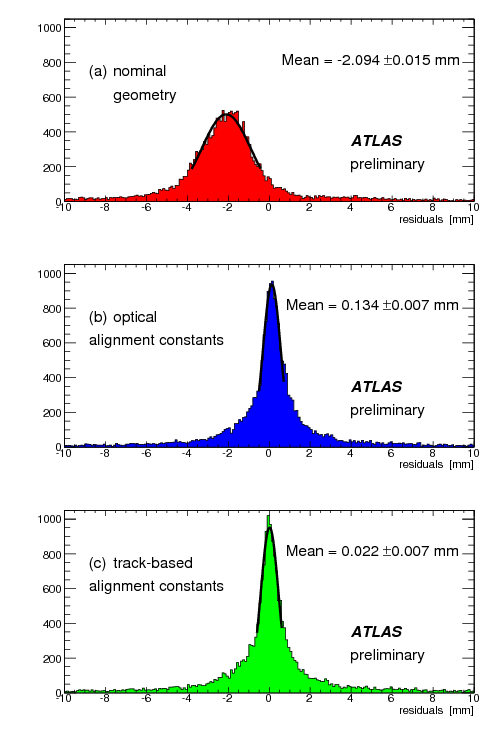}
\caption{Sagitta in barrel cosmic rays with and without alignment
  corrections. Top: with nominal geometry; Middle: with corrections
  from the optical sensors; Bottom: with corrections from straight
  tracks.} \label{fig:barrel_align}
\end{figure}

Figure \ref{fig:barrel_align} shows sagitta measurements for barrel
tracks, with and without alignment corrections.  For the barrel the
alignment corrections were derived from both optical sensors and
straight tracks, since the optical system of the barrel only covers
half of the chambers.  The corrections from straight tracks work
considerably better those from the barrel optical system, due to some
limitations in the optical system which are being addressed by
additional commissioning work.  The alignment with tracks corrects the
sagitta to within 30 $\mu$m as required for optimal spectrometer
performance, whereas the barrel optical system is currently only able
to correct chamber positions to 100-200 $\mu$m.

\section{Conclusions}

The ATLAS muon system is nearly fully installed and tested in the
ATLAS cavern.  Data from cosmic rays show that the detector is
performing well with cosmic rays though a complete evaluation of
detector performance will only be possible with collision data.  The
spectrometer will be ready for the first collisions data from the LHC,
expected in fall 2009.


\begin{thebibliography}{9}   
\bibitem{atlas_paper}   ``The ATLAS Experiment at the CERN Large
  Hadron Collider'', The ATLAS Collaboration, G Aad {\it et al} 2008
  JINST {\bf 3} S08003, http://www.iop.org/EJ/abstract/1748-0221/3/08/S08003
\bibitem{calib_paper} ``Calibration model for the MDT
  chambers of the ATLAS Muon Spectrometer'', P. Bagnaia {\it et al}, ATLAS Note
  ATL-MUON-PUB-2008-004, http://cdsweb.cern.ch/record/1089868
\bibitem{testbeam_paper} ``System Test of the ATLAS Muon Spectrometer
  in the H8 Beam at the CERN SPS'', Adorisio {\it et al}, C
  Nucl. Instrum. Methods Phys. Res., {\bf A593}, 3 (2008) 232-254, http://cdsweb.cern.ch/record/1056267
\end{thebibliography}

\end{document}